\documentclass[%
 aip,
 amsmath,amssymb,
 reprint,%
]{revtex4-1}

\usepackage{graphicx}
\usepackage{color}
\usepackage{dcolumn}
\usepackage{bm}
\usepackage[mathlines]{lineno}
\usepackage[utf8]{inputenc}
\usepackage[T1]{fontenc}
\usepackage{mathptmx}
\usepackage{etoolbox}

\makeatletter
\def\@email#1#2{%
 \endgroup
 \patchcmd{\titleblock@produce}
  {\frontmatter@RRAPformat}
  {\frontmatter@RRAPformat{\produce@RRAP{*#1\href{mailto:#2}{#2}}}\frontmatter@RRAPformat}
  {}{}
}%
\makeatother
\begin{document}

\preprint{AIP/123-QED}

\title{A spatially varying differential equation for multi-patch pandemic propagation}
\author{Abhimanyu Ghosh}%
\affiliation{Poolesville High School, Poolesville, MD}
%

\date{\today}

\begin{abstract}
We develop an extension of the Susceptible-Infected-Recovery (SIR) model to account for spatial variations in population as well as infection and recovery parameters. The equations are derived by taking the continuum limit of discrete interacting patches, and results in a diffusion equation with some nonlinear terms. The resulting population
dynamics can be reinterpreted as a nonlinear heat flow equation where the temperature vector captures both infected and recovered populations across multiple patches. 
\end{abstract}

\maketitle

%

\subsection{\label{sec:level1}Introduction}
The modeling of epidemiological events such as pandemics is important for predicting trends in disease propagation \cite{cov1,cov2,turkyilmazoglu22,turkyilmazoglu23,kroger}. Pandemic prediction is typically split between continuum models, such as differential equation based models, as well as multi-patch models based on discrete hopping events between individual nodes (or patches). Of the two, the former allows simple intuitive solutions that could help with building intuition and policy decisions. In the past, we worked with the susceptible-infected-recovery (SIR) differential equations \cite{kermack1927}, which  resembled models based on semiconductor physics such as the drift-diffusion model \cite{abhi04}. Furthermore, functional parameters of that model (e.g. infection rate) were related to physical parameters such as reproduction number and incubation time. 
However, that model was based on a single patch, i.e, a spatially independent set of infection and recovery parameters. However, in reality these parameters vary over space due to migration of populations and differences in infection rate and mitigation across different regions. Accordingly, a continuum differential equation accounting for spatial variations would be needed to replace multi-patch models in discrete mathematics in order to allow physically meaningful solutions. 

In compartmental models \cite{kendall,reed-frost}, it is not uncommon to include a diffusion term\cite{noble}  $\sim Dd^2I/dx^2$ to account for spatial spread. However, the origin of that diffusion term, or the relation to physical parameters are not commonly discussed. 

In this paper, we derive a generalization to the SIR model that account for spatial variations in the susceptible, infected and recovery variables (S, I, R), as well as the infection-recovery parameters themselves, $A, B$. We show that the derived equations follow a conservation law associated with a population diffusion, related with migration dynamics. The equation is then converted to a single vector variable, and shown to resemble the nonlinear heat conduction equation (where heat diffuses into areas of less heat). In contrast to phenomenological diffusion terms, we derive it by taking a continuum limit of the underlying discrete interacting population dynamics, which gives us a way to relate the diffusion constant to conventional SIR parameters. Furthermore, we find that the dynamics follows a nonlinear diffusion equation of the form $\sim SDd^2I/dx^2$, which makes the underlying solutions richer in dynamics. 
\\\\
\subsection{Recap: A single-patch SIR Model}
Let us recap the parts of the SIR model, and its solutions. The conventional SIR model is 
\begin{eqnarray}
    \frac{dS}{dt} &=& -ASI \nonumber\\
    \frac{dI}{dt} &=& ASI-BI \nonumber\\
    \frac{dR}{dt} &=& BI 
\end{eqnarray}
The population $S + R + I = N$ is fixed in this analysis, as is obvious by adding their time derivatives, since the right hand side vanishes when summed. 
\\\\
In our past paper, we rewrote the equation in scaled format in terms of susceptible and infected fractions $s = S/N_p,~~ f = I/N_p$, 
\begin{eqnarray}
    \frac{ds}{dt} &=& -\dfrac{sf}{\tau_0} \nonumber\\
    \frac{df}{dt} &=& \dfrac{(s-\beta)f}{\tau_0} 
\end{eqnarray}
where $\tau_0 = 1/AN_p$ is the incubation time of the disease, and $\beta = B/AN_p = 1/R_0$ is the mitigation efficiency, with $R_0$ being the reproduction number (how many people each infected person infects on average) and $N_p$ is the population of an infected cluster (or patch), defined as a group whose infected and recovery rates are approximately the same (with some slight fluctuation). 
\\\\
As we argued in our previous paper, when the reproduction number $R_0 < 1$ and $\beta > 1$, the pre-factor of the second equation $(s-\beta)$ is always negative (the fraction $s < 1$), and the fraction of infected decreases from its initial value $f_0 = 1-s_0$ to zero, meaning the infection dies out before it spreads, and thus no pandemic occurs. The corresponding long-time solutions to the above equations are $s^* = s(t\rightarrow \infty) = 1, ~f^* = f(t\rightarrow \infty) = 0$. However, if $\beta = 1/R_0 < 1$, ie, the infection does spread, the long term solution $f^*$ is again zero, but this time there is a residual infected population and the rest recovered, so that the long term solutions are given by
\begin{eqnarray}
    r^* &=& -\beta\ln{(s^*/s_0)}, ~~ s_0 \approx 1\nonumber\\
    s^* &=& -\frac{1}{\beta}W\left(-\frac{e^{1/\beta}}{\beta}\right) \approx \beta^{5/2}
\end{eqnarray}
where $W$ is the Lambert-W function, roughly equal to a power law. 
\\\\
We will now generalize the SIR equation to a spatially varying population.


\subsection{Continuum multi-patch SIR model} 

Let us now generalize the equation, assuming constant $A, B$, but allowing $S, I$ to vary with x, picking up an index $S_x$, $I_x$. For a 1-D grid along x, the $S_x$ at a particular point $x$ is influenced by local infected populations $I_x$ as well as neighboring populations $I_{x\pm \Delta x}$
\begin{eqnarray}
 \ldots &&\ldots \nonumber\\
 \frac{dS_x}{dt} &=& -A_xS_x\Biggl[I_x + I_{x+\Delta x} + I_{x-\Delta x}  \ldots\Biggl]
\end{eqnarray}
For an infinitesimally small grid spacing $\Delta x \rightarrow 0$, assuming only nearest neighbor interactions, we can simplify this as
\begin{equation}
    \frac{dS_x}{dt} \approx -A^\prime_xS_x\frac{d^2I_x}{dx^2} - 3 A_xS_xI_x
\end{equation}
where we used the finite difference representation for a second derivative
\begin{equation}
    \frac{d^2I}{dx^2} \approx
    \frac{I_{x+\Delta x} + I_{x-\Delta x} - 2I_x}{\Delta x^2}
\end{equation}
proportional to the difference between the function $I_x$ and the average $(I_{x+\Delta x} + I_{x-\Delta x})/2$ of its neighbors. Here, $A^\prime_x = A_{x}\Delta x^2$. By extension then, the final equation for the scaled SIR model looks like  
\begin{eqnarray}
\frac{d{s}}{dt} &=& - D s\dfrac{d^2f}{dx^2} - Asf  \nonumber\\
\frac{d{f}}{dt} &=& f\Biggl(D\dfrac{d^2s}{dx^2} + As - \beta\Biggr)
\end{eqnarray}
where the diffusion constant represents the rate of change of positional spread in infection peaks per time interval, and $\beta$ is the recovery parameter
\begin{equation}
  D = A^\prime N_x, ~~~ \beta = B^\prime_x/A^\prime_x N_x 
\end{equation} $D$ resemebles a diffusion terms with units of distance squared per time interval representing the rate of change of variance in $x$, where $N_x$ is the local population, similar to the single patch equation. 
\\\\
The equation makes intuitive sense, as the susceptible population decreases if there is a variation in $I$ with a net influx of infected population, meaning the rate of incoming infected is larger than the outgoing (difference in derivatives). 
If we allow for $D$ to vary with position as well, the equation modifies a bit
\begin{eqnarray}
\frac{d{s}}{dt} &=& -  s\Biggl[\dfrac{d}{dx}\Biggl(D\dfrac{df}{dx}\Biggr)+Af\Biggr]  \nonumber\\
\frac{d{f}}{dt} &=& f\Biggl[\dfrac{d}{dx}\Biggl(D\dfrac{ds}{dx}\Biggr) + As - \beta\Biggr]
\end{eqnarray}
\subsection{Relation to thermal transport}
Let us look at a similar equation, that of heat conduction. The equation for the  temperature variation with position in 1-D looks like
\begin{equation}
    \dfrac{dT}{dt} = K\dfrac{d^2T}{dx^2} + G
\end{equation}
The modified SIR equation has a similar form, except it has two variables that depend on each other in a cross-linked way (the $s$ equation depends on $f$ to the right, and vice-versa), and there is an extra pre-factor $s$ or $f$. Furthermore, the diffusion term has the opposite sign, since interactions with infected populations always reduce $S$, unless we reinsert the recovered population back into the susceptible as in an SIS model. We can absorb the nonlinear term using $d\ln s/dt = (ds/dt)/s$ to give us
\begin{eqnarray}
\frac{d{\ln{s}}}{dt} &=& - D \dfrac{d^2f}{dx^2}  \nonumber\\
\frac{d{\ln{f}}}{dt} &=& \Biggl(D\dfrac{d^2s}{dx^2} - \beta\Biggr)
\end{eqnarray}
Here we are dropping the $\pm Asf$ terms, assuming that we are looking at the deviations from their long-term single patch solutions $f \rightarrow 0$ and $s \rightarrow s^*$.

To resemble the conventional heat conduction equation, we rewrite the equations in matrix form as 
\begin{equation}
    \frac{d}{dt}\left(\begin{array}{c}\ln{s} \\ \ln{f}\end{array}\right) = \left(\begin{array}{cc}0 & -D\\ D & 0\end{array}\right)\dfrac{d^2}{dx^2}\left(\begin{array}{c}{s} \\  {f}\end{array}\right)+  \left(\begin{array}{c}{0} \\  {-D\beta}\end{array}\right)
\end{equation}
which can be written to resemble the heat conduction equation as
\begin{equation}
    \dfrac{d\ln{\bar{T}}}{dt}= {\bar{K}}\dfrac{d^2\bar{T}}{dx^2} + {\bar{G}} 
\end{equation}
where the temperature vector, thermal conductance tensor, and generation vector are respectively
\begin{eqnarray}
    {\bar{T}} &=& \left(\begin{array}{c}s \\ f\end{array}\right) \nonumber\\
    {\bar{K}} &=& \left(\begin{array}{cc}0 & - D \\ D & 0\end{array}\right) \nonumber\\
    {\bar{G}} &=& \left(\begin{array}{c}0 \\ -D\beta\end{array}\right) 
\end{eqnarray}

Note also that in contrast to the heat conduction equation, there is a disparity between the variables to the right and left, with the left involving a logarithm while the right does not. This makes the equation nonlinear. If $T_1$ and $T_2$ are two solutions to the equation, then their linear sum $T_1 + T_2$ will not solve the equation, because $\ln{(T_1 + T_2)}$ is not equal to $\ln{T_1} + \ln{T_2}$ but rather the product $\ln{T_1T_2}$. 

\subsection{Solving the nonlinear heat conduction equation}
\begin{figure}[!ht]
\includegraphics[width=0.5\textwidth]{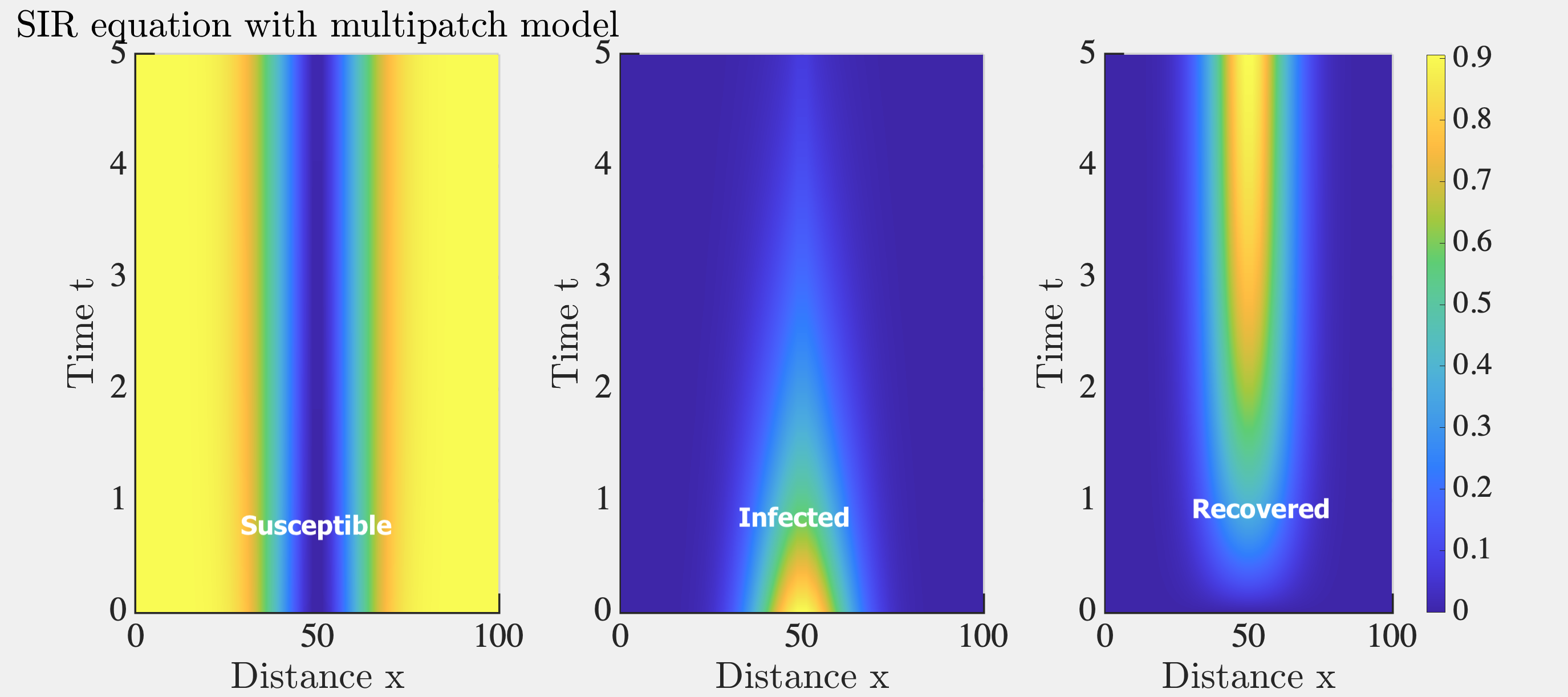}
    \caption{\it{Evolution of fractional susceptible, infected and recovered populations over time, starting with an initial infected peak at a location and then spreading out across the neighboring zones through infection and recovery.}}
\label{patch}
\end{figure}

Fig.~\ref{patch} shows a case study of the heat conduction equation in 1-D to resemble the equivalent S-I-R evolution dynamics over time. The initial condition is a Gaussian distribution of infected population around the middle of our simulation grid (around $x = 50$). Over time, the infected population reduces while the recovered increases as the susceptible population reaches a steady-state solution. We will discuss more details in an upcoming preprint. 



\bibliography{aipsamp}

\end{document}